\documentclass[11pt,oneside]{article}
\usepackage{geometry}
\usepackage{graphicx}
\usepackage{dcolumn}
\usepackage{bm}
\usepackage{setspace}
\usepackage{subfigure}
\newcommand{\car}{$^{12}$C }
\newcommand{\oxy}{$^{16}$O }
\newcommand{\caI}{$^{40}$Ca }
\newcommand{\cer}{{\cal R}}
\newcommand{\cde}{{\cal D}}

\begin{document}
\title{\bf Testing superscaling predictions in \\
electroweak excitations of nuclei}
\author{M. Martini$^{\,1}$, G. Co'$^{\,1}$, 
M. Anguiano$^{\,2}$ and A. M. Lallena$^{\,2}$\\
\small 
1) Dipartimento di Fisica, Universit\`a di Lecce and\\
\small 
 Istituto Nazionale di Fisica Nucleare sez. di Lecce,\\
\small 
 I-73100 Lecce, Italy\\
\small 
2) Departamento de F\'{\i}sica
\small 
 At\'omica, Molecular y Nuclear,\\
\small 
 Universidad de Granada, 
\small 
 E-18071 Granada, Spain}
\maketitle
\begin{abstract}
  Superscaling analysis of electroweak nuclear response functions is
  done for momentum transfer values from 300 to 700 MeV/c.  Some
  effects, absent in the Relativistic Fermi Gas model, where the
  superscaling holds by construction, are considered.  From the
  responses calculated for the $^{12}$C, $^{16}$O and $^{40}$Ca nuclei,
  we have extracted a theoretical universal superscaling function
  similar to that obtained from the experimental responses.
  Theoretical and empirical universal scaling functions have been used
  to calculate electron and neutrino cross sections. These cross
  sections have been compared with those obtained with a complete
  calculation and, for the electron scattering case, with the
  experimental data.
\end{abstract}



\section{Introduction}
The properties of the Relativistic Fermi Gas (RFG) model of the
nucleus \cite{alb88} have inspired the idea of superscaling.  In the
RFG model, the responses of the system to an external perturbation are
related to a universal function of a properly defined scaling variable
which depends upon the energy and the momentum transferred to the
system. The adjective universal means that the scaling function is
independent on the momentum transfer, this is called scaling of first
kind, and it is also independent on the number of nucleons, and this
is indicated as scaling of second kind.  The scaling function can be
defined in such a way to result independent also on the specific type
of external one-body operator.  This feature is usually called scaling
of zeroth-kind \cite{mai02,ama05a,ama05b}.  One has superscaling when
the three kinds of scaling are verified. This happens in the RFG model.

The theoretical hypothesis of superscaling can be empirically tested
by extracting response functions from the experimental cross sections
and by studying their scaling behaviors.  Inclusive electron
scattering data in the quasi-elastic region have been analyzed in this
way \cite{mai02,don99b}. The main result of these studies is that the
longitudinal responses show superscaling behavior.  The situation for
the transverse responses is much more complicated.

The presence of superscaling features in the data is relevant not only
by itself, but also because this property can be used to make
predictions. In effect, from a specific set of longitudinal response
data \cite{jou96}, an empirical scaling function has been extracted
\cite{mai02}, and has been used to obtain neutrino-nucleus cross
sections in the quasi-elastic region \cite{ama05a}. 

We observe that the empirical scaling function is quite different from
that predicted by the RFG model. This indicates the presence of
physics effects not included in the RFG model, but still conserving
the scaling properties.  We have investigated the superscaling
behavior of some of these effects. They are: the finite size of the
system, its collective excitations, the Meson Exchange Currents (MEC)
and the Final State Interactions (FSI).  The inclusion of these
effects produce scaling functions rather similar to the empirical one.
Our theoretical universal scaling functions, $f_{\rm U}^{\rm th}$, and
the empirical one $f_{\rm U}^{\rm ex}$, have been used to predict
electron and neutrino cross sections.

\section{Superscaling beyond RFG model}
\label{sec:res1}
The definitions of the scaling variables and functions, have been
presented in a number of papers \cite{alb88,mai02,ama05a,ama05b}
therefore we do not repeat them here. 
The basic quantities calculated in our work are the electromagnetic,
and the weak, nuclear response functions. We have studied their
scaling properties by direct numerical comparison (for a detailed
analysis see Ref. \cite{mar06}).

\begin{figure}
\begin{center}
\includegraphics[clip,height=16cm]{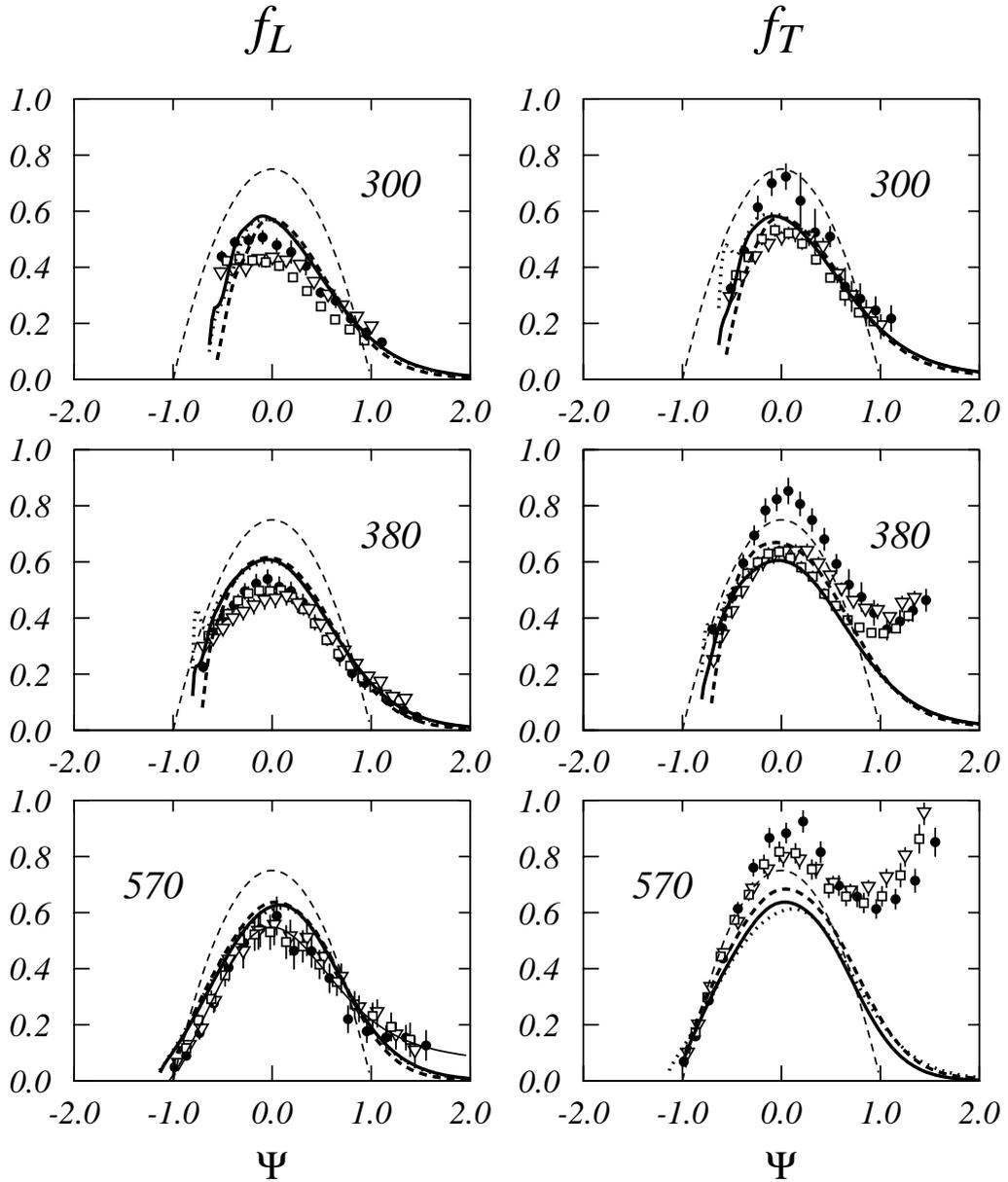}
\end{center}
\caption{\small 
  Empirical longitudinal, $f_L$, and transverse, $f_T$, scaling
  functions obtained from the experimental electromagnetic responses
  of Ref.  \protect\cite{jou96}.  The numbers in the panels indicate the
  values of the momentum transfer in MeV/c. The full circles refer to
  \car, the white squares to \caI, and the white triangles to
  $^{56}$Fe.  The thin black line in the $f_L$ panel at 570 MeV/c, is
  the empirical scaling function obtained from a fit to the data.  The
  thick lines show the results of our calculations when all the
  effects beyond the RFG model have been considered.  The
  full lines have been calculated for \car, the dotted lines for \oxy,
  and the dashed lines for \caI. The dashed thin lines show the RFG
  scaling functions.}
\label{fig:fexp}
\end{figure}

We present in Fig. \ref{fig:fexp} the experimental longitudinal and
transverse scaling function data for the \car, \caI and $^{56}$Fe
nuclei given in Ref. \cite{jou96} for three values of the momentum
transfer. 
We observe that the $f_L$ functions scale better than the $f_T$ ones.
The $f_T$ scaling functions of \car, especially for the lower $q$
values, are remarkably different from those of \caI and $^{56}$Fe.

The observation of the figure, indicates that the scaling of first
kind, independence on the momentum transfer, and of zeroth kind,
independence on the external probe, are not so well fulfilled by the
experimental functions. These observations are in agreement with those
of Refs.  \cite{mai02,don99b}.

To quantify the quality of the scaling between a set of 
$M$ scaling functions, each of them 
known on a grid of $K$ values of the scaling variable
$\Psi$, we define the two indexes:
\begin{equation}
\cde \, = \, \max_{i=1,\ldots,K} 
\left\{ \max_{\alpha=1,\ldots,M} \left[ f_\alpha(\Psi_i) \right]
 \, - \, \min_{\alpha=1,\ldots,M} \left[ f_\alpha(\Psi_i) \right] 
\right\} \, ,
\label{eq:delta} 
\end{equation}
and
\begin{equation}
{\cal R}=\frac {1}{K f^{\max}} \sum_{i=1,\ldots,K} 
\left\{ \max_{\alpha=1,\ldots,M} \left[ f_\alpha(\Psi_i) \right]
 \, - \, \min_{\alpha=1,\ldots,M} \left[ f_\alpha(\Psi_i) \right] 
\right\}
\label{eq:erre}
\end{equation}
where $f^{\rm max}$ is the largest value of the $f_\alpha$.

The two indexes give complementary information.  The $\cde$ index is
related to a local property of the functions: the maximum distance
between the various curves. Since 
the value of this index could be
misleading if the responses have sharp resonances, 
we have also used the $\cer$ index which is instead sensitive to global
properties of the differences between the functions. Since we know
that the functions we want to compare are roughly bell shaped, we have
inserted the factor $1/f^{\rm max}$ to weight more the region of the
maxima of the functions than that of the tails.
\begin{table}[ht]
\begin{center}
{\begin{tabular}{|c|c|c|}
\hline
          & \multicolumn{2}{c|}{$f_{\rm L}$} \\
\hline 
$q$ [MeV/$c$] & $\cde$ & $\cer$  \\
\hline
   300    & 0.107 $\pm$ 0.002 & 0.152 $ \pm$ 0.013 \\ 
   380    & 0.079 $\pm$ 0.003 & 0.075 $ \pm$ 0.009 \\
   570    & {\bf 0.101 $\pm$ 0.009} & {\bf 0.079 $ \pm$ 0.017} \\
\hline
          & \multicolumn{2}{c|}{$f_{\rm T}$} \\  
\hline
   300    &
            0.223 $\pm$ 0.004 &  0.165 $\pm$ 0.017  \\
   380    &
            0.235 $\pm$ 0.005 &  0.155 $\pm$ 0.014    \\
   570    &
            0.169 $\pm$ 0.003 &  0.082 $\pm$ 0.007  \\
\hline
\end{tabular}}
\caption{\label{tab:rdelta}Values of the $\cde$ and $\cer$ 
indexes, for the experimental 
scaling functions of  Fig. \protect\ref{fig:fexp}.}
\end{center}
\end{table}

In Tab. \ref{tab:rdelta} we give the values of the indexes calculated
by comparing the experimental scaling functions of the various nuclei
at fixed value of the momentum transfer. We consider that the scaling
between a set of functions is fulfilled when $\cer <$ 0.096 and $\cde
<$ 0.11. These values have been obtained by adding the uncertainty to
the values of $\cer$ and $\cde$ for $f_L$ at 570 MeV/c.
From a best fit of this last set of data we extracted an empirical
universal scaling function \cite{mar06} represented by the thin full
line in the lowest left panel of Fig. \ref{fig:fexp}.  This curve is
rather similar to the universal empirical function given in Ref.
\cite{mai02}.

Let's consider now the scaling of the theoretical functions.  The thin
dashed lines of Fig. \ref{fig:fexp} show the RFG scaling functions.
The thick lines show the results of our calculations when various
effects beyond the RFG are introduced, \textit{i.e.}: nuclear finite
size, collective excitations, final state interactions, and, in the
case of the $f_T$ functions, meson-exchange currents.

We have studied the effects of the nuclear finite size, by calculating
scaling functions within a continuum shell model. At q=700 MeV/c,
these scaling functions are very similar to those of the RFG model. At
lower values of the momentum transfer, the shell model scaling
functions show sharp peaks, produced by the shell structure, not
present in the RFG model. We found that shell model scaling functions
fulfill the scaling of first kind, the most likely violated, down to
400 MeV/c.

We have estimated the effects of the collective excitations by doing
continuum RPA calculations with two different residual
interactions\cite{bot05}.  The RPA effects become smaller the larger
is the value of the momentum transfer.  At $q >$ 600 MeV/c, the RPA
effects are negligible if calculated with a finite-range interaction.
Collective excitations breaks scaling properties, but we found that
scaling of first kind is satisfied down to about 500 MeV/c.
 
The presence of the MEC violates the scaling of the transverse
responses.  We included the MEC by using the model of Ref.
\cite{ang02}. In our calculations only one-pion exchange diagrams are
considered, including those with virtual excitation of the $\Delta$.
In our model MEC effects start to be relevant for $q \sim$ 600 MeV/c.
We found that MEC do not destroy scaling in the kinematic range of our
interest.

The main modification of the shell model scaling functions, are
produced by the FSI, we have considered by using the model developed
in Ref. \cite{bot05}.  We obtained scaling functions very different
from those predicted by the RFG model, and rather similar to the
empirical ones.  In any case, the FSI do not heavily break the scaling
properties.  We found that the scaling of first kind is conserved down
to $q$=450 MeV/c.

The same type of 
scaling analysis applied to $(\nu_e,e^-)$ reaction leads to
very similar results \cite{mar06}.

\section{Superscaling Predictions}
\label{sec:res2}

To investigate the prediction power of the superscaling hypothesis, we
compared responses, and cross sections, calculated by using RPA, FSI
and eventually MEC, with those obtained by using $f_{\rm U}^{\rm th}$
and $f_{\rm U}^{\rm exp}$.

\begin{figure}
\begin{center}
\includegraphics[clip,height=15cm]{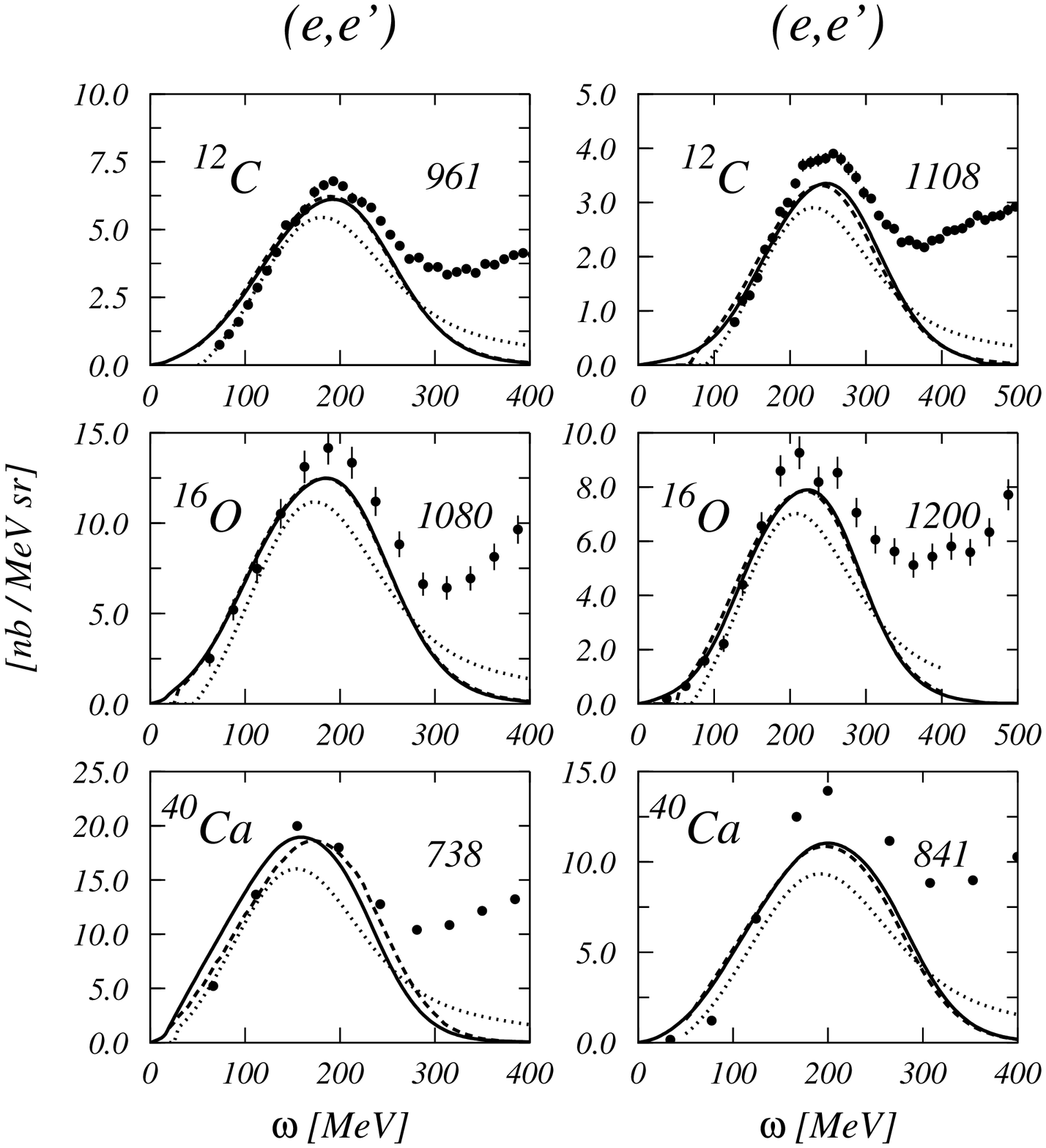}
\end{center}
\caption{\small 
  Inclusive electron scattering cross sections.  The numbers in
  the panels indicate, in MeV, the energy of the the incoming
  electron.  The \car data \protect\cite{sea89} have been measured at
  a scattering angle of $\theta$=37.5$^o$, the \oxy data
  \protect\cite{ang96a} at $\theta$=32.0$^o$ and the \caI data
  \protect\cite{wil97} at $\theta$=45.5$^o$.  The full lines show the
  results of our complete calculations. The cross sections obtained 
  by using $f_{\rm U}^{\rm th}$  are shown by the dashed lines,
  and those obtained with $f_{\rm U}^{\rm ex}$ 
  by the dotted lines.}
\label{fig:ee_xsect}
\end{figure}

We show in Fig. \ref{fig:ee_xsect} double differential electron
scattering cross sections calculated with complete model (full) and
those obtained with $f_{\rm U}^{\rm th}$ (dashed lines) and $f_{\rm
  U}^{\rm exp}$ (dotted lines).  These results are compared with the
data of Refs.  \cite{sea89,ang96a,wil97}.

The excellent agreement between the results of the full calculations
and those obtained by using $f_{\rm U}^{\rm th}$, indicates the
validity of the scaling approach in this kinematic region where the
$q$ values are larger than 500 MeV/c. The differences with the cross
sections obtained by using the empirical scaling functions, reflect
the differences between the various scaling functions shown in Fig.
\ref{fig:fexp}.  The disagreement with the experimental data is
probably due to the fact that our models do not consider the
excitation of the real $\Delta$ resonance, and the pion production
mechanism.
\begin{figure}
\begin{center}
\includegraphics[clip,height=15cm,angle=90]{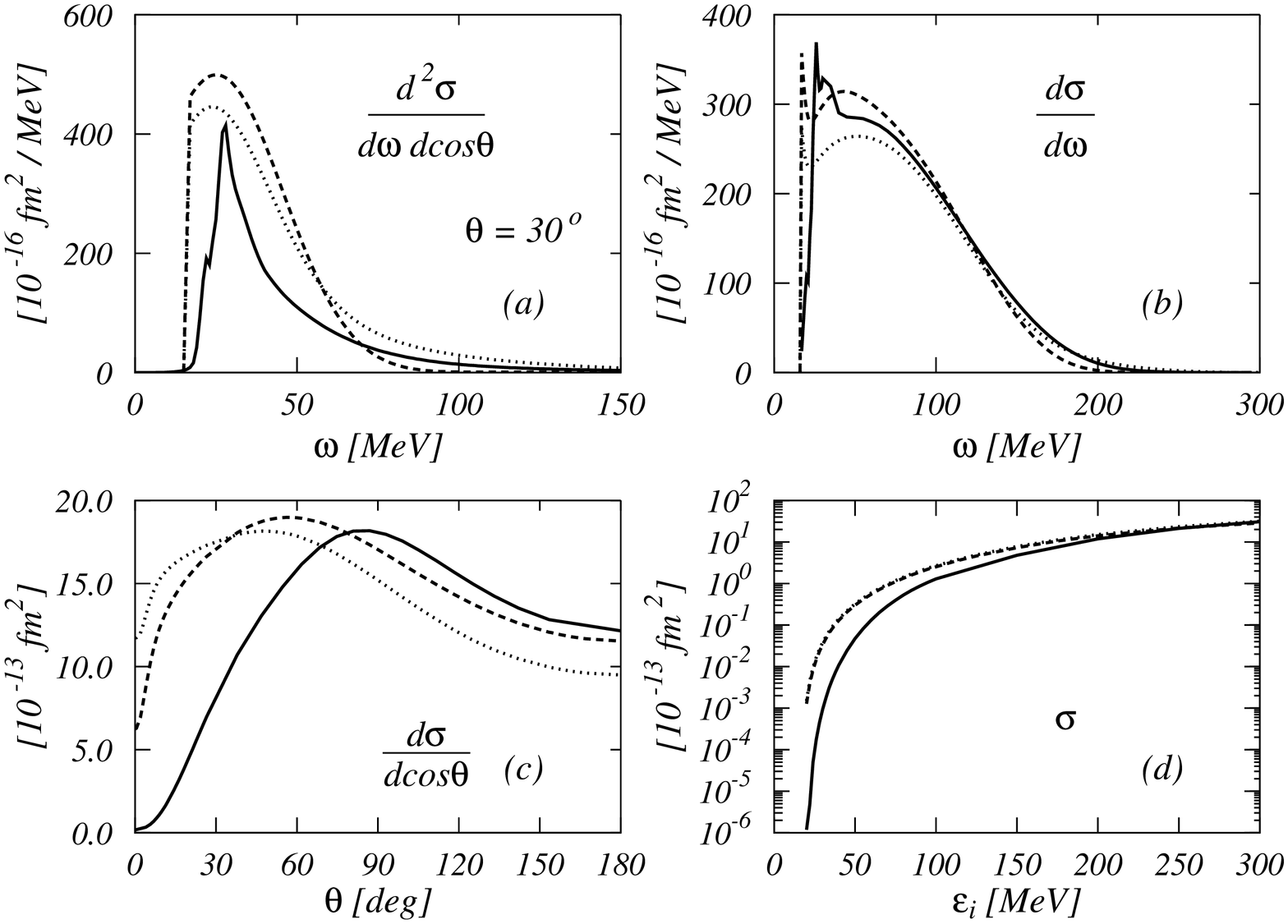}
\end{center}
\caption{\small 
  Neutrino charge exchange cross sections on \oxy.  
  In all the panels the full lines show the result of our
  complete calculation, the dashed (dotted) lines the result
  obtained with our universal (empirical) scaling function.
  The results shown in panels (a), (b) and (c) 
  have been obtained for neutrino energy of 300 MeV.
  Panel (a): double differential cross sections calculated
  for the scattering angle of 30$^o$ as a function of the nuclear
  excitation energy. Panel (b): cross sections
  integrated on the scattering angle, always as a function of the
  nuclear excitation energy. Panel (c): cross sections
  integrated on the nuclear excitation energy, as a function of the
  scattering angle. Panel (d): total cross sections, as a function of 
  the neutrino energy.}
\label{fig:nue}
\end{figure}

The situation for the double differential cross sections is well
controlled, since all the kinematic variables, beam energy, scattering
angle, energy of the detected lepton, are precisely defined, and
consequently also energy and momentum transferred to the target nucleus.
This situation changes for the total cross sections which are of major
interest for the neutrino physics.  The total cross sections are only
function of the energy of the incoming lepton, therefore they consider
all the scattering angles and of the possible values of the energy and
momentum transferred to the nucleus, with the only limitation of the
global energy, and momentum, conservations.  This means that, in the
total cross sections, kinematic situations where the scaling is valid
and also where it is not valid are both present.

We show in the first three panels of Fig.  \ref{fig:nue} various
differential charge-exchange cross sections obtained for 300 MeV
neutrinos on \oxy target. In the panel (a) we show the double
differential cross sections calculated for a scattering angle of
30$^o$, as a function of the nuclear excitation energy. The values of
the momentum transfer vary from about 150 to 200 MeV/c. This is not
the quasi-elastic regime where the scaling is supposed to hold, and
this explains the large differences between the various cross
sections.

The cross sections integrated on the scattering angle are shown as a
function of the nuclear excitation energy in the panel (b) of the
figure, while the cross sections integrated on the excitation energy
as a function of the scattering angle are shown in the panel (c). 
The first three panels of the figure illustrate in different manner the same
physics issue. The calculation with the scaling functions fails in
reproducing the results of the full calculation in the region of low
energy and momentum transfer, where surface and collective effects are
important. This is shown in panel (b) by the bad agreement between the
three curves in the lower energy region, and in panel (c) at low
values of the scattering angle, where the $q$ valued are minimal. 

Total charge-exchange neutrino cross sections are shown in panel (d)
as a function of the neutrino energy $\epsilon_i$. The scaling
predictions for neutrino energies up to 200 MeV are unreliable. These
total cross sections are dominated by the giant resonances, and more
generally by collective nuclear excitation. We have seen that these
effects strongly violate the scaling. At $\epsilon_i$ =200 MeV the
cross section obtained with our universal function is still about 20\%
larger than those obtained with the full calculation.  This difference
becomes smaller with increasing energy and is about the 7\% at
$\epsilon_i$ = 300 MeV. This is an indication that the relative weight
of the non scaling kinematic regions becomes smaller with the
increasing neutrino energy.

\bibliographystyle{ws-procs9x6}

\end{document}